\documentclass[amsmath,amssymb,lengthcheck,onecolumn,aps,prl] {revtex4}
\usepackage[T1]{fontenc}
\usepackage[utf8]{inputenc}
\usepackage{graphics}
\usepackage{graphicx}

\usepackage{xcolor}
\begin{document}
 %\begin{frontmatter}
\title{\textbf{Collective Excitation of Quantum Droplet with Different Ranges of the Interaction of \\ P\"oschl-Teller Potential}}
%Insights into the Collective Excitation Mechanisms in Quantum Droplet via P\"oschl-Teller Potential with Different Interaction Ranges
\author{Avra Banerjee\footnote{2020php001.avra@students.iiests.ac.in,  avrabanerjee1@gmail.com; Corresponding author}}

 \author{Dwipesh Majumder\footnote{dwipesh@physics.iiests.ac.in}}
 
\affiliation{Department of Physics, Indian Institute of Engineering Science and Technology, Shibpur, WB, India}
%\date{\today}

\begin{abstract}
In this article, we studied quantum droplet with the P\"oschl-Teller (PT) interaction potential between the Bose atoms. The Gross-Pitaevskii (GP) equation governs the system. The range and strength of the PT interaction can be adjusted. First, we studied the quantum droplet's density variation for various PT interaction parameters by the imaginary-time split-step Crank-Nicolson (CN) method. We then used the Bogoliubov theory to examine the collective excitation spectra. We observed that sharp roton forms and phonon modes are missing during long-range interactions. There is a gap at the zero momentum zone due to the long-range PT interaction, which increases with the range and strength of the interaction.
\end{abstract}

\maketitle

\section*{Introduction}

The Bose-Einstein condensation (BEC) of an ideal gas serves as the model for quantum statistical phase transitions \cite{einstein}. Owing to the low atomic energies and densities obtained in the experiments, mean-field theory—specifically, the Gross-Pitaevskii (GP) energy functional—may provide a good representation of the underlying structure of a BEC. Even with weak interactions between its atoms, an ultracold gas may show drastically different behavior in the quantum degenerate domain from that of an ideal gas. The only substantial two-body s-wave scattering occurs at low temperatures and densities. 

The liquid form of BEC is known as quantum droplet \cite{Petrov2015,Trarruell2018,ref1}. Studies on the droplet under short and long-range interactions can be found here \cite{Trarruell2018,drop_exp3, Trarruell2018PRL, dipolar_droplets,DDI1,DDI2,DDI6,DDI7,DDI_ref1,DDI1_ref1,ref2}. A proper balance between the mean-field and beyond mean-field interactions is required to form the droplet \cite{LHY}. In addition to this system, the quantum droplet has also been observed as a cigar-shaped droplet in the asymmetric dipolar interacting systems \cite{drop_exp2}. Including three-body interaction, quantum droplets can also be formed \cite{Adhikary}. Using an expanded Gross-Pitaevskii equation (GPE) with the Lee-Huang-Yang (LHY) energy functional $E_{LHY}$, the experimental results can be qualitatively interpreted \cite{tpau}. Since quantum fluctuations are significant, relying on theoretical frameworks that can explain effects other than mean-field corrections is essential. The Bogoliubov theory incorporates perturbative fluctuations at the lowest order for homogeneous systems in the thermodynamic limit and gives a sufficient description in the weakly interacting limit \cite{montecarlo}. Quantum droplet can be a combination of atoms from two distinct elements  \cite{PRL89, PRL100,Itali2020} or atoms with two different internal degrees of freedom of the same element \cite{spin_BEC, PRL'101}.

The dipole-dipole interaction (DDI) may substantially affect the dynamics, excitations, and thermodynamic properties of the BEC \cite{DDI11,DDI12}. New quantum phases like droplet and supersolid states are also generated \cite{supersolid,supersolid1,supersolid2}. The pair potential also describes the Coulomb interaction, which is a long-range interaction \cite{coloumb5,coloumb3,coloumb1}. In the absence of contact interactions and DDI, infrared-divergent components are exactly eliminated in the Hartree-Fock-Bogoliubov (HFB) shift of the single particle excitation energy \cite{coloumb6}. In addition to dipolar and coulomb interactions, another long-range potential called PT interactions between atoms has been studied \cite{PT,PT2,PT_ref1,PT1}.

The motivation for studying the PT interaction potential between Bose atoms is as follows: There is a parameter $\mu$ that controls the range of interaction, which is a unique and interesting feature of this potential. With a single type of potential, one can study the effects of both long- and short-range interactions. We have studied and observed when and how roton minima developed for different ranges and strengths ($U$) of the PT interaction. Also, long-range interaction creates a gap \cite{gap,gap1} in the zero-momentum region, which depends upon the range and strengths of PT interaction. The slope of the excitation spectrum, which often divides into two sections to allow for collective excitations of phonon and roton types, is a well-known method for determining the critical velocity of the superfluid \cite{coll1,coll2}. Roton exhibits a parabolic dispersion at a finite momentum \cite{roton1,roton2}. Bragg spectroscopy has been used to detect phonon and roton structures \cite{roton3,roton4}.

We aim to study how the collective excitation of quantum droplet depends upon PT interaction between Bose atoms. Double-species BECs in the liquid phase have been used in this study. The Gross-Pitaevskii (GP) energy functional, a component of mean-field theory, provides a good approximation of the underlying structure of a droplet. The quantum droplet is self-bound with uniform density. We studied what happens to the droplet when it is perturbed and used the Bogoliubov method to study the collective excitation. The Bogoliubov theory makes it easy to look into collective excitation in a uniform system. In our study, we have considered a very large droplet, so the excitation inside the droplet is not affected by the surface.

\section*{Model and calculation}

\begin{figure}

\includegraphics[width=0.45\textwidth] {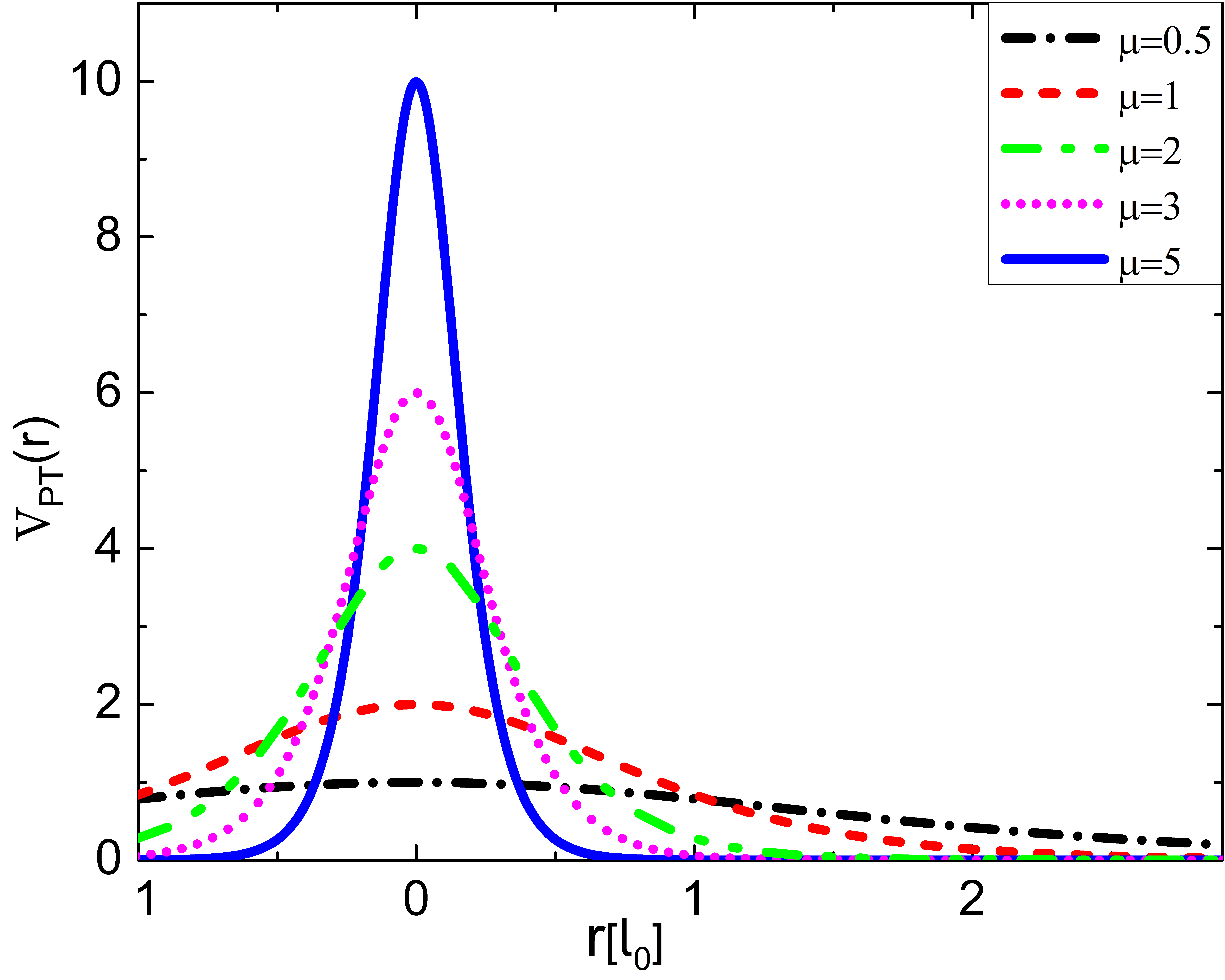}
\caption{  P\"oschl-Teller interaction potential, $V_{PT}$ as a function of the separation distance of two particles in units of $l_0$ for different values of $\mu$ and $U=1$. We plotted the potential on both sides to see the delta function nature of the interaction, though the value of $r$ is solely positive. As we increase the value of $\mu$, the potential becomes more and more like delta-function potential.\\}
\label{pt_uni}
\end{figure}

 We have considered P\"oschl-Teller interaction \cite{PT,PT1}  between dilute Bose atoms in addition to contact interaction. The PT interaction potential ($V_{PT}$) is given by

\begin{equation}
V_{PT}=U\sum_{i<j} \frac{2\mu}{cosh^2(\mu r_{ij})}  
\end{equation}

where $U$ is the strength of the interaction, $r_{ij}$ is the separation between the two particles, and $\mu$ is the interaction parameter expressed as the breadth of the interaction (i.e., $1/\mu$). Various $\mu$ have been considered to explore the excitation's dependent nature. As we increase the $\mu$'s value, its potential nature changes, as seen in Figure (\ref{pt_uni}). The benefit of this model potential is that it allows us to switch from a short-range ($\delta$ function) to a long-range interaction, altering the parameter ($\mu$).

\begin{figure}

\includegraphics[width=0.45\textwidth]  {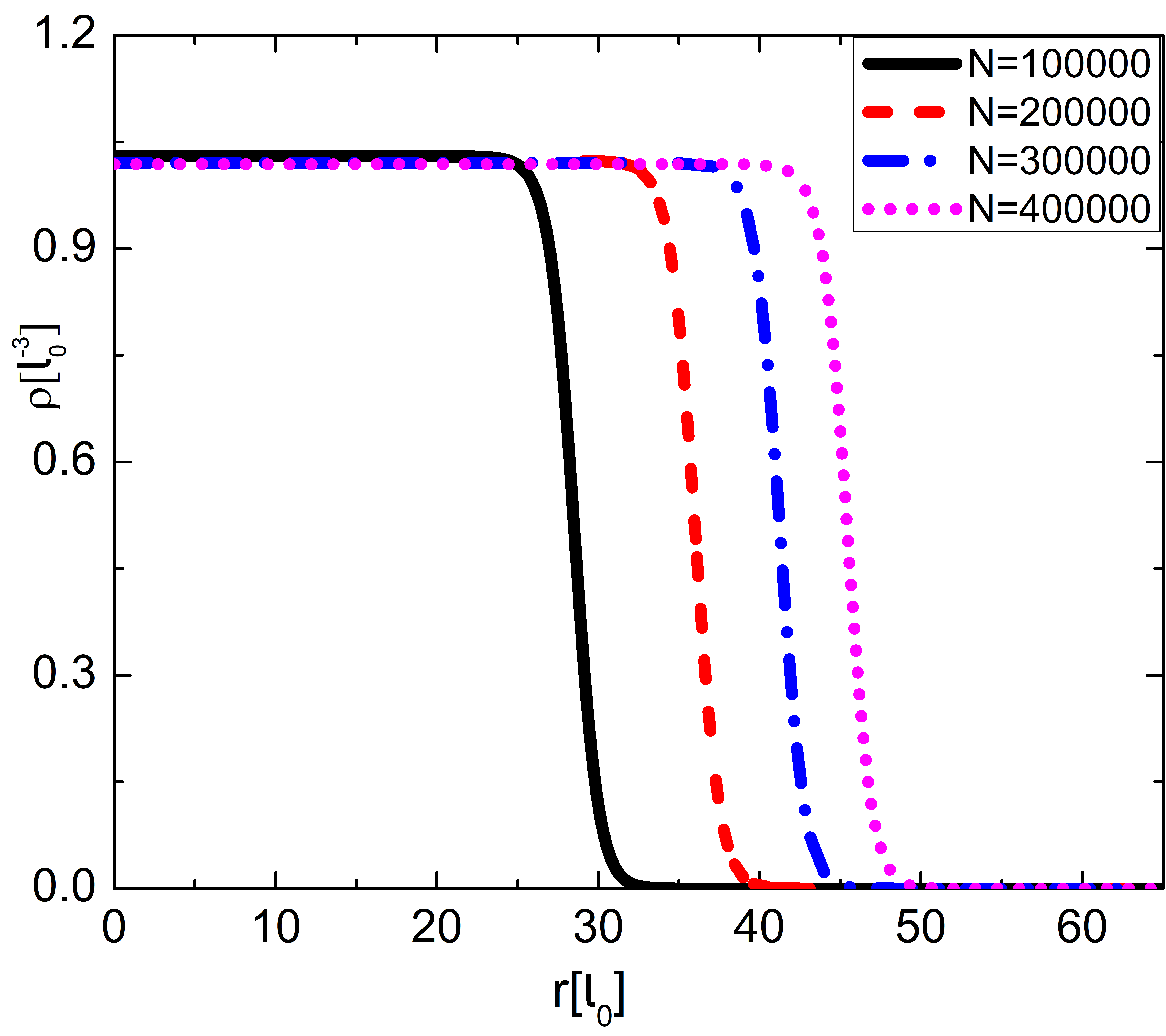}
  \caption{Density variation for different values of particle numbers $(N)$ (without PT potential) shows that the density is uniform and independent of the number of particles. So, one can say that density will be approximately equivalent for an infinitely large system when $N \to \infty $.}
\label{N_uni}
 \end{figure}

We consider a mixture of two species of Bose atoms (two different internal degrees of freedom of the same isotope of an element, ($m_1=m_2=m$) \cite{spin_BEC, PRL'101}, we have used this to simplify the droplet system as we aim to study the effect of PT interaction between atoms) interacting via PT potential. $V(\vec{r}_1,\vec{r}_2) = g_{ij} \delta(\vec{r}_1-\vec{r}_2)$ is the contact interaction between two Bose atoms at ultracold temperature and low density, where $g_{ij}$ is the interaction strength between the atoms of $i^{th}$ and $j^{th}$ species, which may be written as $g_{ij}=4\pi \hbar^2 a_{ij}/m$, where $m$ is the condensed atom's mass and $a_{ij}$ denotes s-wave scattering length, which may be controlled by Feshbach resonance \cite{fesh,fesh1,fesh2}. Near the phase transition, the mean-field GP equations are not enough to study the nature of the droplet; we need to consider the self-repulsive beyond-mean field, Lee-Huang-Yang (LHY) \cite{LHY,LHY1} term (quantum fluctuation term, $g_{LHY}$ is the strength of the quantum fluctuation).\\  The GP equation for the quantum droplet (denoted by $\psi_1$ for one species and $\psi_2$ for another species) can be expressed as ($i=1,2$ and $\texttt{i}=\sqrt{-1}$) \cite{ref10,adhikary_dip}

\begin{widetext}
\begin{eqnarray}
\texttt{i} \frac{\partial \psi_i}{\partial t}=   [-\frac{\nabla  ^2}{2}+  g_{ii}|\psi_i|^2-g_{12}|\psi_{(3-i)}|^2  +g_{LHY}|\psi_i|^3   + \int d\textbf{r}'V_{PT}(\textbf{r}-\textbf{r}')\rho(\textbf{r}',t)]\psi_i
\label{gp_eq}
\end{eqnarray}\\
\end{widetext}

\begin{figure}

  \includegraphics[width=0.46\textwidth]  {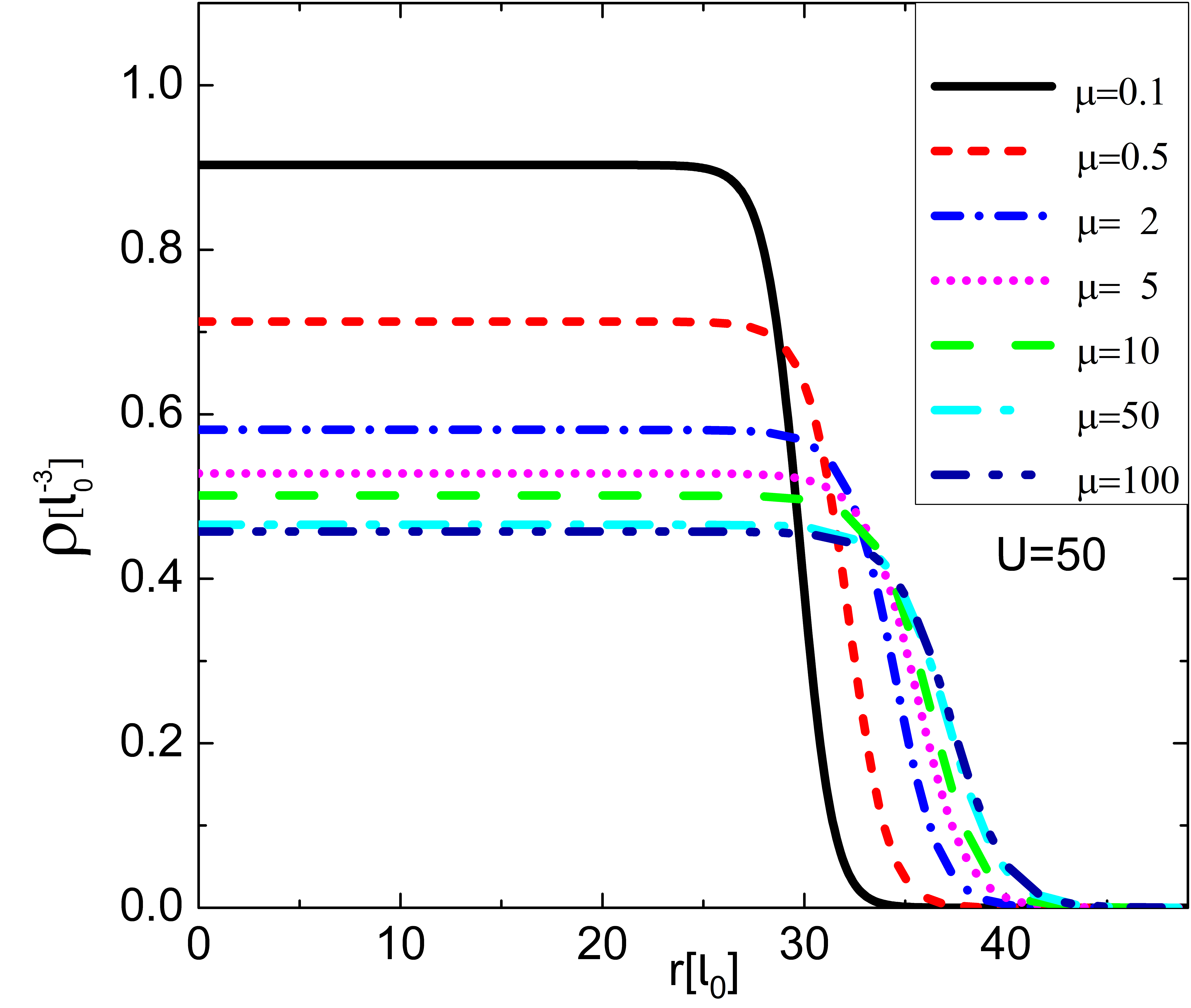}
  \caption{   Density variation for different values of $\mu$ and $U = 50$. As $\mu$ increases, the droplet's density tends to saturate to a value as the PT potential takes on the characteristics of a delta function. The number of particles is $N=100000$; the density variation is equivalent for a higher number of particles ($N$), only the radial extension increases with $N$ as shown in Fig. (\ref{N_uni}).}
\label{PT_drop}

 \end{figure}
 
 \begin{figure}

  \includegraphics[width=0.48\textwidth]  {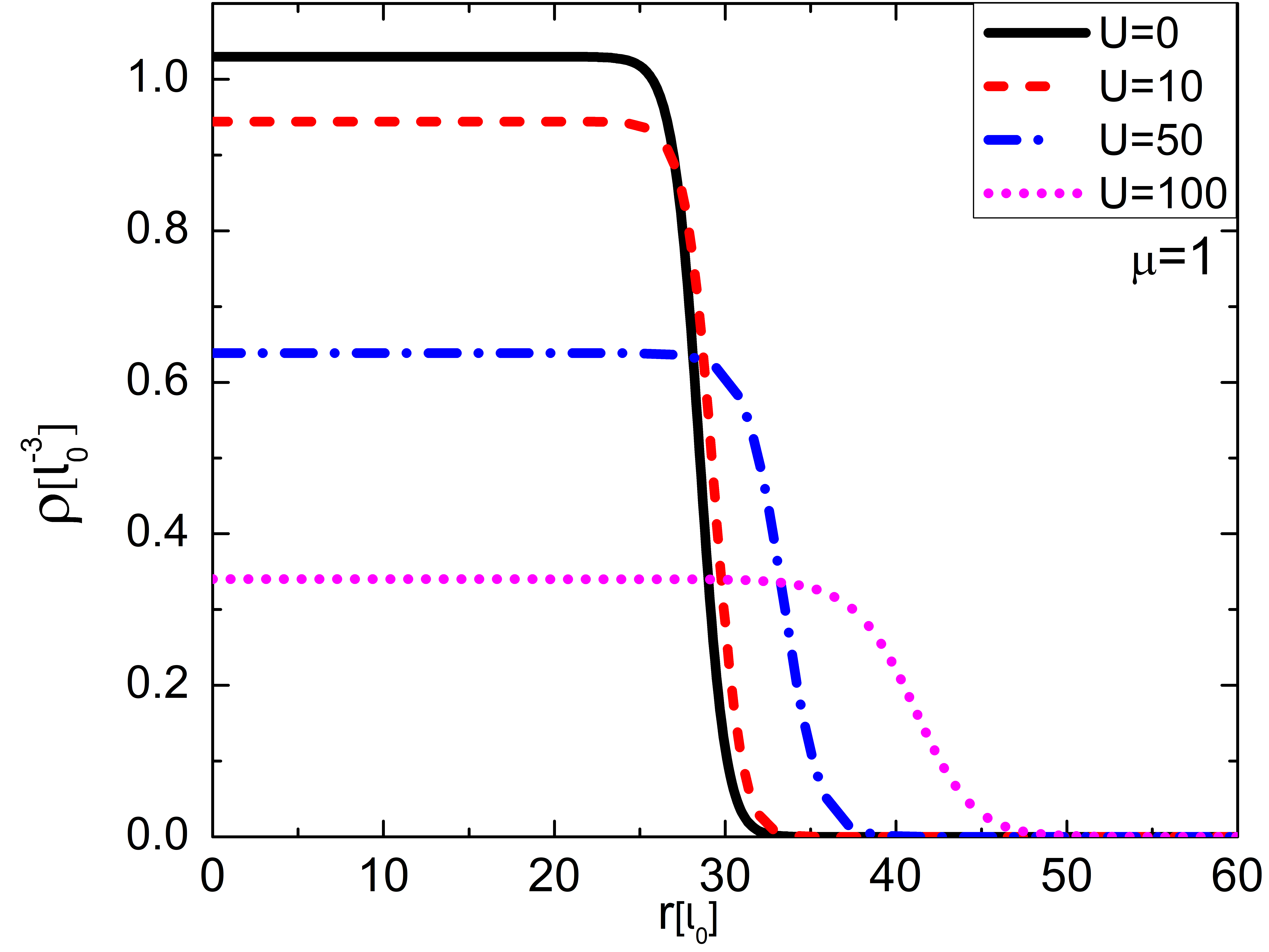}
  \caption{Density variation for different values of $U$ and $\mu$= 1. The strength of the PT interaction potential increases with increasing $U$. Hence, the droplet's density drops as $U $ rises. $U = 0$ for the PT interaction-free droplet. The number of particles is $N=100000$; the density variation is equivalent for a higher number of particles ($N$), only the radial extension increases with $N$ as shown in Fig. (\ref{N_uni}).}
\label{den_u}

 \end{figure}
 
Where $\psi = (\psi_1 \psi_2)^T$ is the two-component wave function and $\rho=|\psi_1|^2+|\psi_2|^2$. $1^{st}$ term on the right side of the GP equation represents the kinetic energy; $2^{nd}$ term is the weak repulsive contact interaction ($g_{ii}=g=4\pi a$ where $a$ denotes scattering length of the interaction) between the same species of atoms; $3^{rd}$ term represents the attractive interaction between two different species of atoms; and $4^{th}$ term is the quantum correction (LHY term). The last integration term is for the PT interaction potential between the atoms. The length scale used in the calculations (of the order of coherence length of BEC \cite{length}) is $l_{0}=1 \mu m$, time and energy in $\frac{ml_0^2}{\hbar}$ and $\frac{\hbar^{2}}{ml_{0}^{2}}$ units, respectively. The condensed atom's mass is $m$. Equal number of particles for each species is considered.

The integral can be simplified in Fourier space using convolution as follows, and it is performed using the Fast Fourier transformation program \cite{adhikary_dip}.
\begin{small}
\begin{eqnarray}
\int d\textbf{r}' V_{PT}(\textbf{r}-\textbf{r}')\rho(\textbf{r}',t) =\int \frac{d\textbf{k}}{(2\pi)^3}e^{-\texttt{i} \textbf{k.r} } \tilde{V}_{PT}(\textbf{k})\tilde{\rho}(\textbf{k},t)
\end{eqnarray}

\end{small}

The Fourier transform of the PT potential is calculated numerically. The Fourier transformation (FT) is defined by \cite{adhikary_dip}

\begin{figure*}

  \includegraphics[width=0.86\textwidth]  {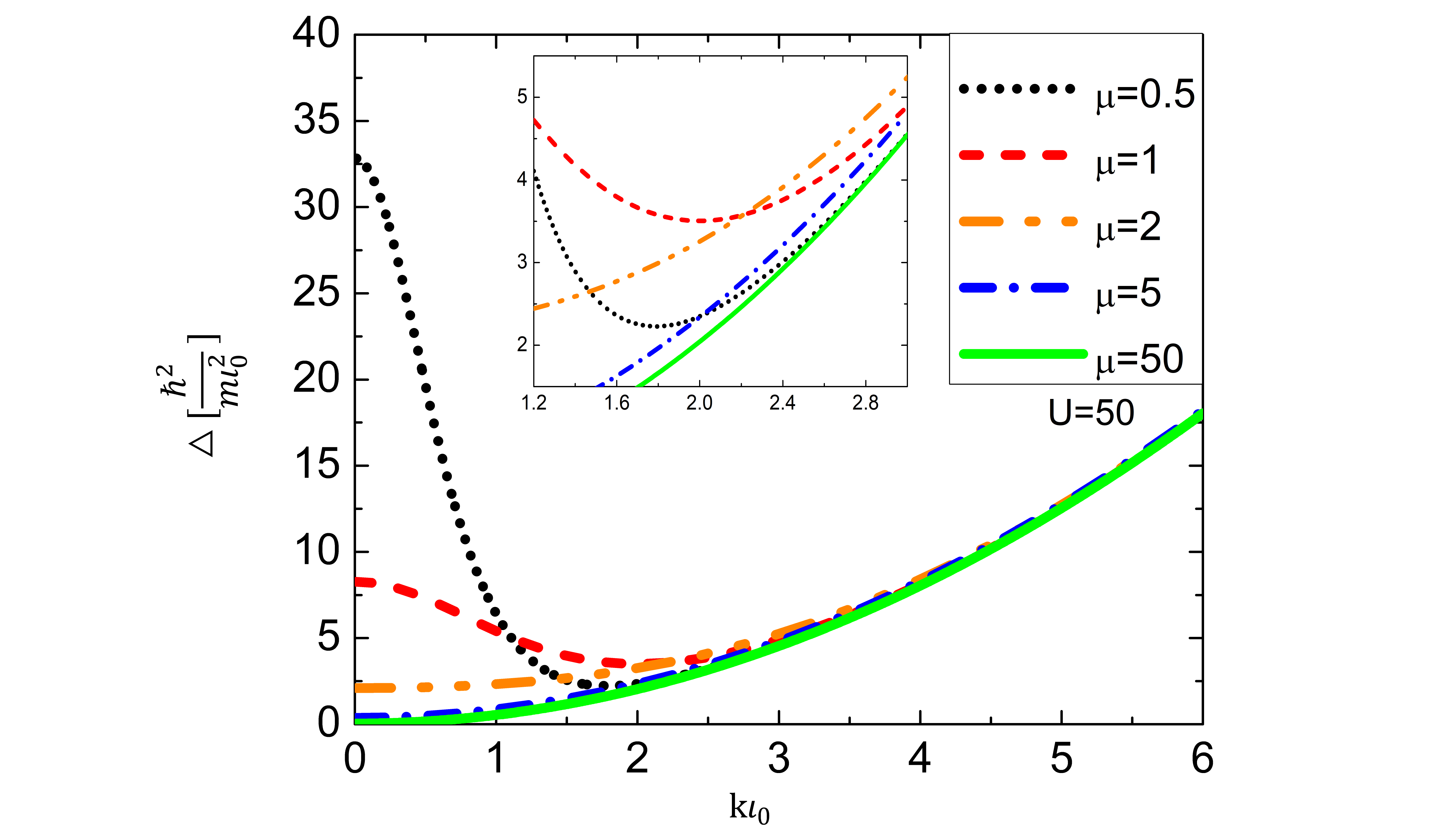}
  \caption{Collective excitation for various $\mu$ and $U=50$. Rotons are observed for the lower values of $\mu$ (phonon mode is missing). A short-range interaction is indicated by a greater $\mu$. At high $\mu$ values, roton starts to disappear. The roton moves towards the higher momentum region with less sharpness for larger $\mu$. At zero momentum ( $k = 0$), a gap \cite{gap,gap1} appears caused by the long-range  PT interaction potential. With higher $\mu$, phonon mode appears due to short-range interaction. Zoomed roton zones are also added on the upper side.}
\label{ex_mu}

 \end{figure*}
\begin{equation}
\tilde{V}_{PT}(\textbf{k}) =\int d\textbf{r}V_{PT}(\textbf{r})e^{\texttt{i} \textbf{k.r} }%, F(\textbf{r}) =\frac{1}{(2\pi)^3}\int d\textbf{q} \tilde{F}(\textbf{q})e^{-\texttt{i} \textbf{q.r} }
 \end{equation}
 
 As our system is radially symmetric (i.e., spherically symmetric quantum droplet), we chose $\nabla^2$ in spherical coordinates. $\psi_i$ is expressed as $\psi_i = \phi_i(r) /r$.
We have used the imaginary-time split-step Crank Nicolson method to solve the GP equation  \cite{adhikary_cn}  where the $\phi_i(r)$ vanishes at $r=0$ and $r=\infty$. The normalization condition is given by

  \begin{eqnarray}
   4\pi  \int_0^\infty \biggr (|\phi_1|^{2}+|\phi_2|^{2}\biggr )  dr= N
  \end{eqnarray} 
 
where the total number of particles in the droplet is denoted by $N$. In the absence of the harmonic trapping potential and in the thermodynamic limit, the surface effect of the droplet has a negligible contribution (the surface energy ($\omega_s$) is proportional to $N^{-1/2}$, for large $N$, $\omega_s$ is negligible), and we have the uniform solution \cite{hui}.

\section{Collective excitation}

Here, we have used the Bogoliubov approach to get the excitation of the large uniform droplet. The collectively excited state can be written as
 $\psi_i^{\mbox{exc}} = \psi_i+\delta \psi_i$. The fluctuation part is given by \cite{dm_coll, ch_bose},

\begin{normalsize}
\begin{equation}
  \delta \psi_i = e^{-\texttt{i}\mu_s t} \left (U_i e^{\texttt{i}\textbf{k} \cdot  \textbf{r} -\texttt{i}\omega t} + V_i^* e^{-\texttt{i}\textbf{k} \cdot  \textbf{r} +\texttt{i}\omega t}\right )
\end{equation}
\end{normalsize}

where $\mu_s$ is the chemical potential of the system and $\{U_i, V_i\}$ are the excitation amplitudes. We will have the equations for $\{U_i, V_i\}$ (considering the first-order approximation of $\{\delta\psi_i\}$) if we use the excited state in our GP equations (\ref{gp_eq}).

\begin{widetext}

\begin{large}
      \begin{eqnarray}
\left(
    \begin{tabular}{c c c c}
      $H_1+C$ & $-4\psi_1 \psi_2^*$ & $A \psi_1^2$&$ -4\psi_1 \psi_2$\\
       $-4 \psi_1^* \psi_2$ & $H_2+C$ & $-4\psi_1 \psi_2$ &  $B \psi_2^2$\\
       $-A\psi_1^{*2}$ & $4\psi_1^* \psi_2^*$ &$-H_1-C$ & $4\psi_1^* \psi_2$\\
       $4 \psi_1^* \psi_2^*$ & $-B \psi_2^{*2}$ & $4 \psi_1 \psi_2^*$ & $-H_2-C$\\
    \end{tabular}
\right)
\left(
\begin{tabular}{c}
$U_1$ \\$U_2$\\$V_1$\\$V_2$ 
\end{tabular}
\right) = \omega \left(
\begin{tabular}{c}
$U_1$ \\$U_2$\\$V_1$\\$V_2$ 
\end{tabular}
\right) 
\end{eqnarray}
   \end{large}

  \end{widetext}
matrix elements are defined as
\begin{eqnarray}
H_1 &=&  \frac{k^2} {2}+ 2|\psi_1|^2- \mu_s- 4 |\psi_2|^2+\frac{25}{4}|\psi_1|^3 \nonumber \\
H_2 &=&  \frac{k^2} {2}+ 2|\psi_2|^2-\mu_s- 4 |\psi_1|^2+\frac{25}{4}|\psi_2|^3 \nonumber \\
A &=& 1+\frac{15}{4}|\psi_1|  \nonumber \\
B &=& 1+\frac{15}{4}|\psi_2|  \nonumber \\
C &=&  \tilde{V}_{PT}(k) \rho  \nonumber
\end{eqnarray}

     \begin{figure*}

  \includegraphics[width=0.86\textwidth]  {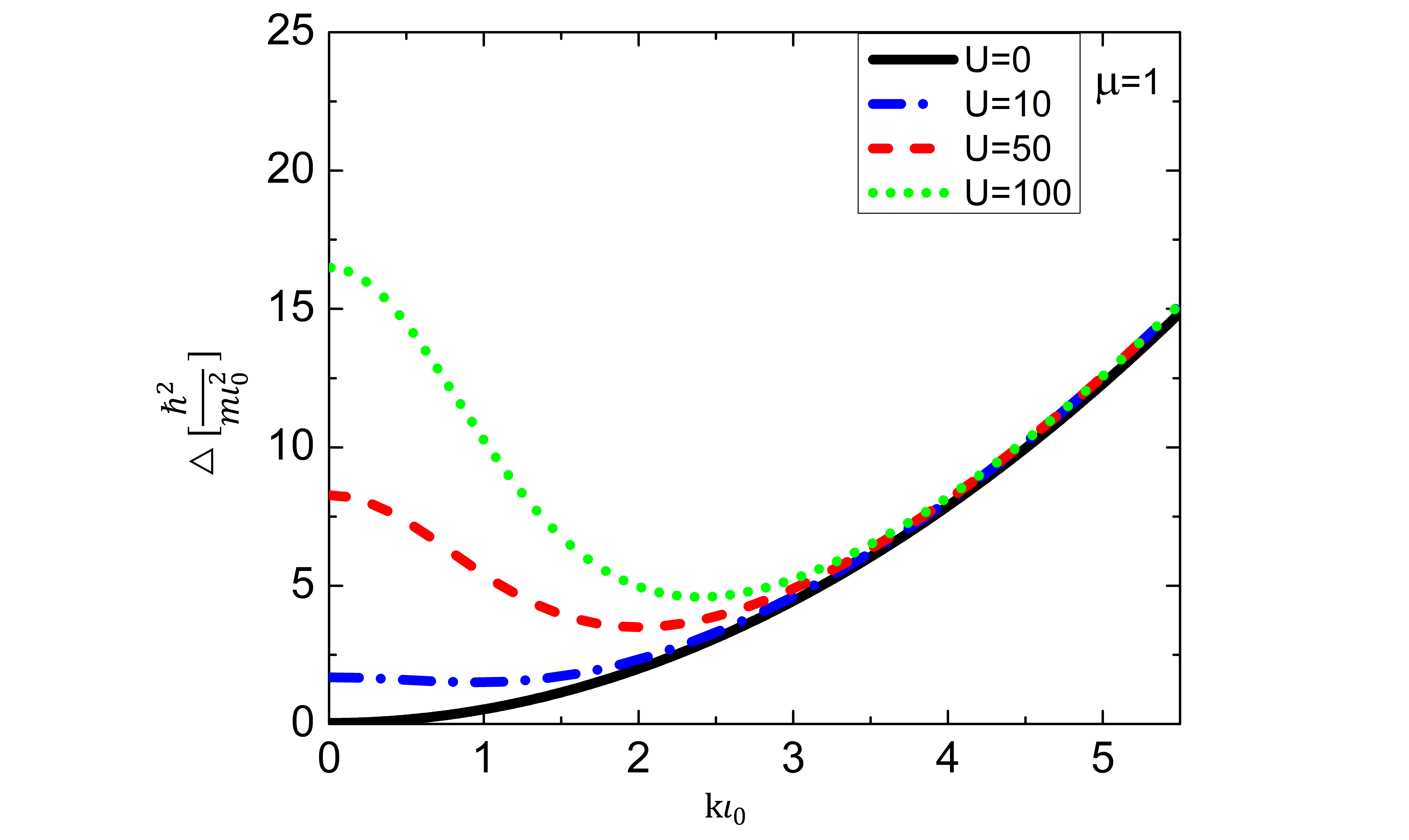}
  \caption{ Collective excitation for different $U$ and $\mu = 1$ values. There is no long-range interaction when $U = 0$, as indicated by the solid black curve. There are just phononic mode spectra as a result. The long-range PT interaction becomes relevant for higher values of $U$. For greater values of $U$, roton therefore begins to form. It is evident that minima become sharper and move towards the higher momentum region as $U$ increases. In addition, a gap \cite{gap,gap1} exists at zero momentum ( $k = 0$) region, which results from long-range PT interaction and increases with the strength of PT interaction.}
\label{ex_u}
\centering
 \end{figure*}
We get the collective excitation spectrum ($\Delta$), shown in Figs. \ref{ex_mu} and \ref{ex_u}, by diagonalizing the above matrix using LAPACK code \cite{lapack}.

\section{Results and discussion}

This study used the P\"oschl-Teller (PT) interaction potential between diluted Bose atoms in addition to contact interaction. Figure (\ref{pt_uni}) shows the PT potential's characteristics. For large values of $\mu$, it is seen that this potential acts like a delta function, which is a contact interaction between atoms. Furthermore, for smaller values of $\mu$, this potential is similar to a long-range potential. The interaction's strength is set by the parameter $U$. The ground state density of the droplet of Bose atoms, without PT interaction, for a range of values of $N$ (number of particles) is shown in Figure (\ref{N_uni}). It has been observed that density is uniform and independent of particle number.

Next, over a range of values of $\mu$, the ground-state density has been studied (Figure (\ref{PT_drop})) by solving the GP equation (\ref{gp_eq}) (by the imaginary-time split-step Crank-Nicolson (CN) method). We have used the values of $g_{12}=4$, $g_{ii}=1$, and $g_{LHY}=5/2$ to get a uniform ground-state density (refer to . \cite{Petrov2015}). In 3-dimensional analysis, $g_{12}$ > $g_{ii}$; otherwise, no droplet will be formed. When $\mu$ increases, density first decreases; nevertheless, density tends to saturate at a value when $\mu$ increases to a sufficiently large value. The larger $\mu$ serves as a contact interaction. For different values of $U$, the ground-state density is displayed in Figure (\ref{den_u}). The droplet's density decreases with increasing $U$, and the stronger the PT interaction strength, the more density decreases. The density variation would be equivalent for more particles as the density of the droplet is independent of the number of particles ($N$); only the radial extension increases with $N$ as shown in Fig. (\ref{N_uni}).

Then, we computed the large uniform quantum droplet's collective excitation using the Bogoliubov method. Excitation energy is shown in the Figs. (\ref{ex_mu}) and (\ref{ex_u}). Figure (\ref{ex_mu}) displays the graphical variations of collective excitation for various values of $\mu$. The range of the interaction potential is long when $\mu$ is lower. For this reason, the roton begins to manifest, and phonon mode is missing. When the interaction becomes short-range, roton vanishes for greater values of $\mu$. The solid green curve shows this is fully phononic at $\mu = 50$. The roton moves towards the higher momentum region with less sharpness for larger $\mu$. Figure (\ref{ex_u}) displays the energy spectra for various values of $U$ and $\mu=1$. In the absence of PT interaction, the spectra have phonon mode of excitation but do not contain any roton mode (solid black line). It is evident that minima become sharper and move towards the higher momentum region as $U$ increases. The long-range character of the PT potential also accounts for the gap at the zero momentum zone ($k = 0$). The gap \cite{gap,gap1,coloumb6} increases with the strength and range of the PT interaction.

In conclusion, we have studied the density variation that saturates to a value when the PT interaction becomes sufficiently short-ranged. The PT interactions have a distinct effect compared to other types of long-range interactions (i.e., dipole-dipole and coulomb interactions), and there is no singularity in the PT potential. In the excitation spectrum, we found that the phonon mode stiffens \cite{gap1} with a stronger and longer range of interactions, whereas studies on dipolar bose gas suggest there are no significant changes in the phonon mode with the dipolar interaction strengths \cite{dip_conclu}, and for the charged bose gas ref. \cite{coulomb_conclu}, the sound velocity increases (i.e., phonon mode energy increases) with increasing coulomb interaction for fixed dipolar strengths. In contrast to their study \cite{dip_conclu, coulomb_conclu}, we also found an interaction range and strength-dependent gap in the lower momentum region ($k = 0$).

We found that the roton in the spectra became more prominent with increasing strengths and ranges, whereas studies on dipolar gas suggest that the roton mode stiffens with increasing dipolar strengths, and the increasing coulomb interaction also stiffens the roton \cite{dip_conclu, coulomb_conclu}. The roton mode appears due to long-range interaction and disappears when PT interaction tends to become a delta function-type interaction.


\begin{thebibliography}{100} 


\bibitem{einstein} A. Einstein, Sitzungsber. Kgl. Preuss. Akad. Wiss, {\bf 261}  (1924); S. N. Bose, Z. Phys. {\bf 26}, 178 (1924).
\bibitem{ref1} A. Khan, A. Debnath, Frontiers in Physics {\bf 10}, 887338 (2022).


\bibitem{Petrov2015} D. S. Petrov, Phys. Rev. Lett. {\bf 115}, 155302 (2015).
\bibitem{Trarruell2018} C. R. Cabrera, L. Tanzi, J. Sanz, B. Naylor, P. Thomas, P. Cheiney, L. Tarruell, Science {\bf 359}, 301 (2018).
\bibitem{drop_exp3} H. Kadau, M. Schmitt, M. Wenzel, C. Wink, T. Maier, I. F.Barbut, T. Pfau, Nat. Phys. {\bf 530}, 194 (2016); I. F. Barbut, H. Kadau, M. Schmitt, M. Wenzel, T. Pfau, Phys. Rev. Lett.  {\bf 116}, 215301 (2016).
\bibitem{DDI2} L .D. Carr, D. DeMille, R.V. Krems, J. Ye, New. J.Phys {\bf  11}, 055049 (2009).
\bibitem{DDI6} M. Lu et al., Phys. Rev. Lett. {\bf  107}, 190401 (2011).
\bibitem{Trarruell2018PRL} P. Cheiney, C. R. Cabrera, J. Sanz, B. Naylor, L. Tanzi, L. Tarruell, Phys. Rev. Lett. {\bf 120}, 135301 (2018). 
\bibitem{dipolar_droplets} Observation of dipolar droplets in 2016: H. Kadau, M. Schmitt, M. Wenzel, C. Wink, T. Maier, I. F. Barbut, T. Pfau, Nat. Phys. {\bf 530}, 194 (2016); 
I. F. Barbut, H. Kadau, M. Schmitt, M. Wenzel, T. Pfau, Phys. Rev. Lett. {\bf 116}, 215301 (2016); I. F. Barbut, M. Schmitt, M. Wenzel, H. Kadau,T. Pfau, J. Phys. B {\bf 49}, 214004 (2016);  M. Schmitt, M. Wenzel, B. Bottcher, I. F. Barbut, T. Pfau, Nat. Phys. {\bf 539}, 259 (2016);  L. Chomaz, S. Baier, D. Petter, M. J. Mark, F. Wachtler, L.  Santos,  F.  Ferlaino,  Phys.  Rev.  X {\bf 6},  041039 (2016).
\bibitem{DDI7} K. Aikawa et al., Phys. Rev. Lett. {\bf  108}, 210401 (2012).
\bibitem{DDI_ref1}  P. B. Blakie, Photonics {\bf 10(4)}, 393 (2023).
\bibitem{DDI1}T. Lahaye et al., Rep. Prog. Phys. {\bf  72}, 126401 (2009).


\bibitem{DDI1_ref1} Au-Chen Lee, D. Baillie, P. B. Blakie, Phys. Rev. Research {\bf 3}, 013283 (2021).
\bibitem{ref2}A. R. P. Lima,  A. Pelster, Phys. Rev. A {\bf 84}, 041604(R) (2011).


\bibitem{LHY} T. D. Lee, C. N. Yang, Phys. Rev. {\bf 105}, 1119 (1957); T. D. Lee, Kerson Huang, C. N. Yang, Phys. Rev. {\bf 106}, 1135 (1957).
\bibitem{LHY1}  A. R. P. Lima,  A. Pelster,  Phys. Rev. A {\bf 86}, 063609 (2012). 
\bibitem{drop_exp2} G. Semeghini, G. Ferioli, L. Masi, C. Mazzinghi, L. Wolswijk, F. Minardi, M. Modugno, G. Modugno, M. Inguscio, M. Fattori, Phys. Rev. Lett. {\bf 120}, 235301 (2018).
\bibitem{Adhikary} S. K. Adhikari, Phys. Rev. A {\bf 95}, 023606 (2017).
\bibitem{tpau}   F. Bottcher, J. N. Schmidt, J. Hertkorn, K. S. H. Ng, S. D. Graham, M. Guo, T. Langen, T. Pfau, Rep. Prog. Phys. {\bf 84} 012403 (2021). 
\bibitem{montecarlo} L. Parisi, G. E. Astrakharchik, S. Giorgini, Phys. Rev. Lett. {\bf  122}, 105302 (2019).
\bibitem{PRL89} G. Modugno, M. Modugno, F. Riboli, G. Roati, M. Inguscio, Phys. Rev. Lett. {\bf 89}, 190404 (2002).
\bibitem{PRL100} G. Thalhammer, G. Barontini, L. De Sarlo, J. Catani, F. Minardi, M. Inguscio, Phys. Rev. Lett. {\bf 100}, 210402 (2008).
\bibitem{Itali2020} A. Burchianti, C. D. Errico, M. Prevedelli, F. Ancilotto, M. Modugno, L. Salasnich, F. Minardi, C. Fort, Condens. Matter { \bf  5}, 21 (2020).
\bibitem{spin_BEC} J. Stenger, S. Inouye, D. M. Stamper-Kurn, H. J. Miesner, A. P. Chikkatur, W. Ketterle, Nat. Phys. {\bf 396}, 345 (1998); M. S. Chang, C. D. Hamley, M. D. Barrett, J. A. Sauer, K. M. Fortier, W. Zhang, L. You, M. S. Chapman, Phys. Rev. Lett. {\bf 92}, 140403 (2004).
\bibitem{PRL'101} S. B. Papp, J. M. Pino, C. E. Wieman, Phys. Rev. Lett. {\bf 101}, 040402 (2008).
\bibitem{DDI11} L.D. Carr, D. DeMille, R.V. Krems,J. Ye, New. J. Phys { \bf 11}, 055049 (2009).
\bibitem{DDI12}  M. A. Baranov, Physics Reports { \bf 464}, 71 (2008).
\bibitem{supersolid}Z.-H. Luo, W. Pang, B. Liu, Y.-Y. Li, B. A. Malomed, Frontiers of Physics { \bf 16}, 32201 (2021).
\bibitem{supersolid1} F. Boëttcher, J.-N. Schmidt, J. Hertkorn, K. Ng, Graham, M. Guo, T. Langen, T. Pfau, Reports on Progress in Physics { \bf 84}, 012403 (2021).
\bibitem{supersolid2} M. Guo, T. Pfau, Frontiers of Physics {\bf  16}, 32202 (2021).
\bibitem{coloumb5} H. Lieb, J. P. Solovej, Commun. Math. Phys. {\bf  217}, 127 (2001).
\bibitem{coloumb3} M. Tamaddonpur, H. Yavari, Z. Saeidi, Low Temp. Phys. {\bf  45}, 1187 (2019).
\bibitem{coloumb1} E. Darsheshdar, H. Yavari, S. M. Moniri, Eur. Phys. J. Plus {\bf  131,} 178 (2016).
\bibitem{coloumb6} B. Davoudi, M. P. Tosi, Phys. Rev. B {\bf  72}, 134520 (2005).

\bibitem{PT1} D. Das, S. Sahu, D. Majumder, Physica B: Condensed Matter   {\bf   550}, 96 (2018).
\bibitem{PT} A. J. Morris, P. L. Rios, R. J. Needs, Phys. Rev. A { \bf 81}, 033619 (2010).
\bibitem{PT2} L. C. Pereira , V. A. d. Nascimento, Materials { \bf 13(10)}, 2236 (2020).
\bibitem{PT_ref1} A. Debnath, A. Khan, Eur. Phys. J. D { \bf 74}, 184 (2020).

\bibitem{gap} A. Boudjemaa, Phys. Lett. A {\bf 465}, 128712 (2023).
\bibitem{gap1} H. Lyu, Y. Zhang, T. Busch, Phys. Rev. A {\bf 106}, 013302 (2022).


\bibitem{coll1} L. Landau,  J. Phys. USSR {\bf  5}, 71 (1941).
\bibitem{coll2} L. Landau,  J. Phys. USSR {\bf  11}, 91 (1947).
\bibitem{roton1} D. Allum, P. Mcclintock, A. Phillips,  R. Bowley, Philos. Trans. R. Soc. London, Ser. A { \bf 284}, 179 (1977).
\bibitem{roton2} C. A. M. Castelijns, K. F. Coates, A. M. Guénault, S. G. Mussett,  G. R. Pickett, Phys. Rev. Lett. { \bf 56}, 69 (1986).
\bibitem{roton3}S.-C. Ji, L. Zhang, X.-T. Xu, Z. Wu, Y. Deng, S. Chen, J.-W. Pan, Phys. Rev. Lett. { \bf 114}, 105301 (2015).
\bibitem{roton4}M. A. Khamehchi, Y. Zhang, C. Hamner, T. Busch, P. Engels,  Phys. Rev. A { \bf 90}, 063624 (2014).
%\bibitem{coloumb4} J. P. Hague, P. E. Kornilovitch, J. H. Samson, A. S. Alexandrov, Phys. Rev. Lett. {\bf  98}, 037002 (2007).
%\bibitem{coloumb2} S. M. Moniri, H. Yavari, E. Darsheshdar, Eur. Phys. J. Plus {\bf  131}, 122 (2016).
\bibitem{fesh} W.-X. Li, Y.-D. Chen, Y.-T. Sun, S. Tung, Paul S. Julienne, Phys. Rev. A {\bf  106}, 023317 (2022).
\bibitem{fesh1} C. Chin, R. Grimm, P. Julienne, E. Tiesinga, Rev. Mod. Phys. {\bf  82}, 1225 (2010).
\bibitem{fesh2} V. Barbe, A. Ciamei, B. Pasquiou, L. Reichsollner, F. Schreck, P. S. Zuchowski, L.M. Hutson, Nature Physics {\bf  14}, 881 (2018).
\bibitem{ref10} Y. V. Kartashov, B. A. Malomed, L. Tarruell, L. Torner, Phys. Rev. A {\bf 98}, 013612 (2018).
\bibitem{adhikary_dip} R. K. Kumar, L. E. Young-S., D. Vudragović, A. Balaž, P. Muruganandam, S.K. Adhikari, Comput. Phys. Commun.  { \bf 195}, 117 (2015) . 
\bibitem{adhikary_cn} P. Muruganandam, S. K. Adhikari,	Comput. Phys. Commun. { \bf 180}, 1888 (2009) . 
\bibitem{hui} H. Hu, X. J. Liu, Phys. Rev.  A {\bf 102}, 053303 (2020).
\bibitem{dm_coll}  A. Banerjee, D. Majumder, J. Low Temp. Phys. {\bf 215}, 64 (2024). 
\bibitem{ch_bose} A. Boudjemaa,  Phys. Lett. A {\bf 465}, 128712 (2023).
\bibitem{length} S. Gautam, A. K. Adhikari, J. Phys. B {\bf 52}, 055302 (2019); Annals of Phys. {\bf 409}, 167917 (2019).
\bibitem{lapack} Matrix diagonalization subroutine is available here (Linear Algebra PACKage) https://www.netlib.org/.

\bibitem{dip_conclu} M. Edmonds, T. Bland, N. Parker, J. Phys. Commun. {\bf 4}, 125008 (2020).
\bibitem{coulomb_conclu} S. M. Moniri, H. Yavari, E. Darsheshdar, Annals of Phys. {\bf 438}, 168788 (2022).
\end{thebibliography}
\end{document}